# A Two-Stage Coordinative Zonal Volt/VAR Control Scheme for Distribution Systems with High Inverter-based Resources

Asmaa Alrushoud, *Student Member,* IEEE, and Ning Lu, *Fellow,* IEEE

*Abstract*--This paper presents a two-stage zonal Volt/VAR control scheme for coordinating inverter-based resources (IBR) with utility-owned voltage regulators (VR) to regulate voltage in unbalanced 3-phase distribution systems. First, correlations between nodal voltages are derived from nodal voltage sensitivity studies. Then, the feeder is partitioned into non-overlapping, weakly-coupled voltage control zones based on nodal voltage correlations. IBR are used in the first stage to regulate voltage changes continuously and VR are used in the second stage to regulate large voltage deviations. An online VR voltage setpoint tuning strategy is developed to reduce excessive tap changes and avoid large voltage fluctuations without retrofitting existing VR controllers. In addition, the proposed algorithm uses real-time voltage measurements only from the critical nodes (typically less than 4% of total nodes) to reduce the sensing and communication needs. Actual distribution feeder topologies and load and PV time-series data are used to verify the performance of the algorithm. Because the method is a rule-based approach, it runs extremely fast, requires fewer measurements, and requires no retrofit to the existing VR control mechanisms. Simulation results show that the performance of the proposed method in terms of voltage control results and average numbers of VR tap changes are satisfactory.

*Index Terms*-- *Coordinative control, critical nodes, distribution systems, inverter-based resources, PV integration, reactive power support, volt/var control, voltage regulation.*

## I. INTRODUCTION

IN power distribution system, conventional Volt/VAR control (VVC) relies mainly on utility-owned voltage control devices, such as capacitor banks, on-load tap changers (OLTC), and step-type voltage regulators (SVRs). In recent years, distribution systems with high-penetration of inverter-based resources (IBRs), primarily photovoltaic (PV) systems, have seen a dramatic increase in the number of switching of Volt-VAR devices, significant shortening their equipment lifespan [1]. Although IBRs' variable power outputs can cause larger, more frequent voltage fluctuations than passive loads, the superb real and reactive power regulating capability of IBRs also makes them potential high-quality VVC resources. However, because the distributed nature of IBRs, it is essential that a coordinative VVC scheme is developed so that they can work with conventional VVC devices in harmony to maintain nodal voltages within operational limits and reduce losses.

A common approach for distribution system VCC is to formulate an optimal reactive power flow problem that is solved either centrally or distributedly [2-7]. A centralized approach assumes that a centralized controller has full observability and controllability of all controllable devices (including utility-owned VVC device and IBRs). Thus, by determining the on/off status or computing the optimal voltage setpoints of those devices, the optimization objectives, such as minimizing voltage fluctuations, losses, the number of switching, or operation costs can be achieved.

Although the centralized VVC has the advantage of achieving global optimum when controlling voltage within the operational limits at both the local and system levels, it relies on the availability of high-quality, wide-band 2-way communication networks and wide deployment of sensor/actuator on IBRs and utility VVC devices, making large-scale implementations cost prohibitive.

Because voltage issues can be best resolved at the place where a voltage violation occurs, optimization-based, zonal VVC approaches draw increasing attention in recent years. The zone-based method approach first partitions a distribution network into non-overlapping zones and then decomposes the VVC optimization problem into sub-problems, each optimizing the control actions inside a VVC zone. In [8-13], a number of partition methods are introduced, including K-means clustering, spectral clustering, affinity propagation clustering, community detection, binary particle swarm optimization (BPSO), and power flow tracing and agglomerative algorithms. In [14, 15], the authors break a distribution feeder into several zones based on the service area and operation reach of a regulating device to reduce computational complexity and minimize VVC devices switching actions.

However, the aforementioned optimization-based zonal VVC approaches still require real-time observability of the whole distribution feeder using state estimation. The main challenge in distribution state estimation is that the lack of redundant, synchronized, and statistical characterized real-time data in current distribution feeders [16]. Another main drawback in the existing approach (e.g., [12, 14-15]) is that the optimization algorithm yields the optimal tap-position instead of the voltage set-point plus a bandwidth used for controlling

---

an SVR. This adds retrofit costs in field implementation. Thus, although the optimization-based control approach can achieve satisfactory results on small-scale actual distribution system [8-9, 12] and/or small IEEE test feeders [9-13], its implementation cost would be high in distribution systems with hundreds or thousands of nodes supplying unbalanced loads with high-penetration of IBRs.

Therefore, in this paper, we developed a two-stage, coordinative zonal VVC algorithm to operate utility-owned VCC devices in conjunction with IBRs. First, correlations between nodal voltages are derived from nodal voltage sensitivity studies. Then, the feeder is partitioned into non-overlapping, weakly-coupled voltage control zones based on nodal voltage correlations. IBRs are used in the first stage to regulate voltage changes continuously and VRs are used in the second stage to regulate large voltage deviations. An online VR voltage setpoint tuning strategy is developed to reduce excessive tap changes and avoid large voltage fluctuations without retrofitting existing VR controllers. In addition, the algorithm uses real-time voltage measurements from only the critical nodes (typically less than 4% of total nodes). This significantly reduces the sensing and communication needs.

The contributions of the paper are summarized as follows.
- Developed a critical node identification method to minimize the monitoring and communication needs.
- Developed a coordinative VVC control algorithm that uses IBRs to cope with fast, quick voltage variations while using conventional VVC devices to cope with larger voltage excursions using measurements only from the critical nodes.
- Treated each phase separately to effectively account for the unbalanced loads in distribution systems.
- Developed a computation-lite, priority-list based zonal IBR control scheme for correcting voltage violations in each IBR zone.
- Used VVC control signals provided by conventional VVC devices (i.e., voltage setpoint, bandwidth and time delays) so no retrofit is required.
- Developed an online voltage setpoint tuning strategy to adjust their voltage set-points of the VVC devices so they can ride through voltage fluctuations caused by IBRs while maintaining voltage within the desired operational limits.

## II. Conventional VCC Control Mechanism

Conventional voltage regulators (VR), such as SVR and OLTC, are autotransformers equipped with a load tap changing mechanism for regulating voltage at their secondary-sides [17]. A typical tap changer can provide a total regulation range of ±10% of the nominal voltage with 16 steps up and 16 steps down. This results in 0.625% change per step or 0.75 V change per step on a 120 V voltage base.

As shown in Fig. 1, there are three adjustable control settings: voltage reference setpoint, $V_{set}^{VR}$, deadband, $DB^{VR}$, and time delay, $t_d^{VR}$ [18]. The secondary voltage of a voltage regulator (VR), $V_{VR,2}$ needs to be maintained within

$$V_{set}^{VR} - \frac{DB^{VR}}{2} \leq V_{VR,2} \leq V_{set}^{VR} + \frac{DB^{VR}}{2} \quad (1)$$

Normally in a long radial distribution feeder, an OLTC is located at the feeder head, usually inside a primary substation. A few SVRs are deployed in a cascade fashion to regulation voltage along the feeder. Coordination between the OLTC and the cascaded SVRs is achieved by time coordination [14, 15, 19], i.e. by setting up different $t_d$ to ensure that VVC actions are taken sequentially. In the example given in Fig.1, we set $t_d^{OLTC} < t_d^{SVR1} < t_d^{SVR2}$.

SVR has a load drop compensation (LDC) feature for regulating voltage at a remote load center node instead of regulating the SVR secondary voltage. Although we did not use LDC to develop our method in this paper, the method can be readily extended to include LDC.

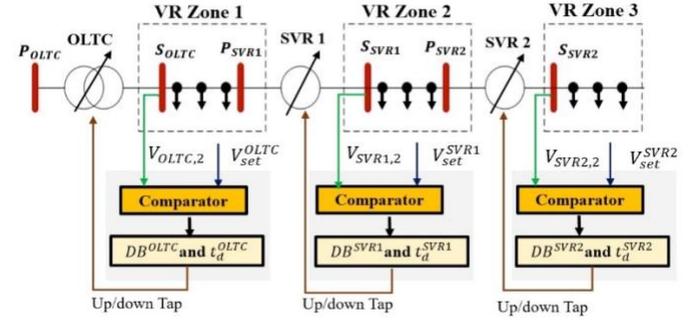

Fig. 1. Simplified distribution feeder.

Conventional VVC devices are tuned and operated assuming one-directional power flow, i.e. voltage drops are one-directional. However, IBRs can inject power into the distribution feeder, causing voltage to increase in the IBR nodes. Thus, even if the secondary voltage of a SVR is within range, the voltages of the downstream nodes may exceed the voltage operation. As shown in Fig. 2, the settings of a VR are: $V_{set} = 122\ V$ and $DB = 4\ V$. When there is no PV in the downstream nodes, all nodal voltages are within the operation limits; when there are PV installed on the downstream nodes, voltages at the end nodes will exceed the upper operation limit. In the next section, we will introduce the 2-stage coordinative VVC for managing the voltage violations in a time-coordination manner that exploits the fast reactive power regulating capability of the IBRs and allow the IBR-based VVC to be seamlessly integrated into the existing VVC control framework.

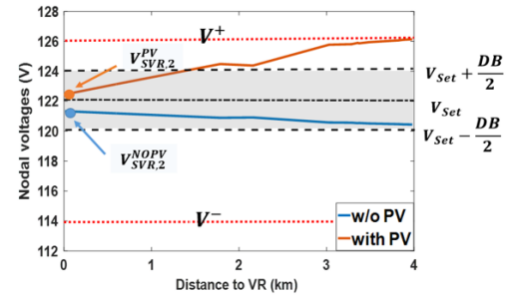

Fig.2. Voltage profile with and without PV downstream an SVR.

## III. CONTROL METHODOLOGY

In this section, we will first give an overview of the two-stage, coordinative zonal Volt/VAR control scheme. Then, the critical node identification method and the first- and second-stage control algorithms are presented.

### A. An Overview of the Algorithm

As shown in Fig. 3, to correct voltage violations locally, the feeder will be partitioned into $K$ IBR control zones ($K = 5$ in Fig. 3a) and $G$ VR control zones ($G = 4$ in Fig. 3b). There is one IBR controller inside each IBR control zone. The controller receives voltage measurements from the critical nodes its control zone. There is one VR controller inside each VR control zone. The VR controller can be installed on a VR device (i.e., a SVR or an OLTC). The VR controller receives measurements from the VR device and the critical nodes inside its zone.

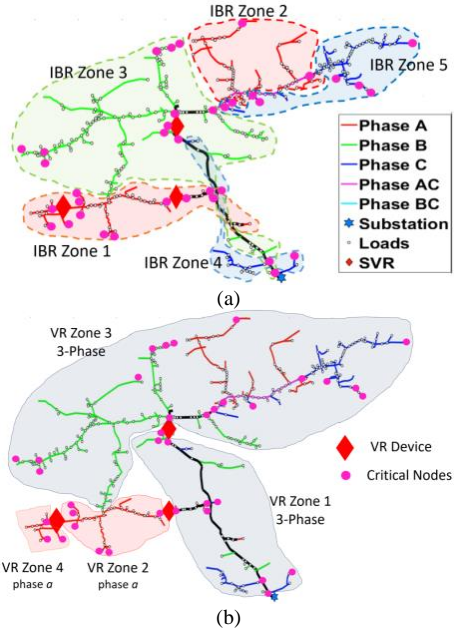

(a)

(b)

Fig. 3. (a) IBR Control Zones and (b) VR Control Zones.

The time coordination between the IBR and VR zones is shown in Fig. 4. In the first stage, a priority-list based reactive power dispatch algorithm is executed with a time step of 1-minute (i.e., $\Delta t = 1$ minute) to decide which IBR in the IBR zone will provide the required reactive power support. The first-stage control objective is to continuously regulate the local voltage violations using the superb reactive power regulating capability of IBRs. In the second stage, the VR controller determines the voltage set-point (i.e., $V_{Set}$) for the SVR or the OLTC at a time step of 2-minute. The second-stage control objective is to remove voltage deviations that cannot be compensated by IBRs.

The time coordination between the upstream and downstream VR devices are done by the time delay settings, as shown in the zoom-in plot in Fig. 4. The time-coordination of the two-stage coordinative VVC exploits the fast reactive power regulating capability of IBRs to effectively reduce the number of switching of VR devices without depleting the IBR reactive power regulation capability.

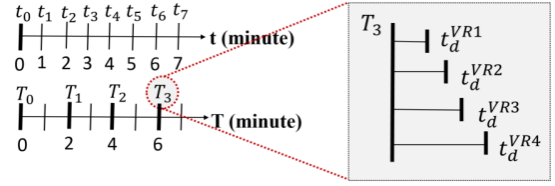

Fig. 4. Time Coordination between IBR and VR zones.

### B. Critical Nodes Identification

To reduce the monitoring and communication needs, we identify a group of critical nodes (see nodes highlighted in purple in Fig. 3), where extreme voltages (i.e., the nodes with the highest or lowest voltages in each zone) are most likely to be observed. There are two approaches for critical nodes identification: statistic-based and model-based. The statistic-based approach can be used if voltage measurements at all load nodes on a distribution feeder are available. If nodal voltage measurements are not available, a modeling based approach is needed to identify the critical nodes.

In this paper, we present a model-based approach using Monte-Carlo simulation. First, we disaggregate the yearly feeder-head profile (normally available to utilities) to each load node using the method introduced by Wang *et. al* in [20]. Then, generate yearly PV outputs from solar irradiance data in the area. After that, a yearly quasi-static power flow (365 days at 1-minute interval) will be run on the test feeder with the targeting PV penetration to obtain the hourly feeder voltage distribution. This allows us to identify the hours where voltage violations are most likely to occur. In the simulation, we consider the PV penetration as the percentage of the peak of the aggregated PV output to the peak load on the feeder.

As shown in Fig. 5, simulation results show that voltage violations are most likely to happen between 9 a.m. and 8 p.m.. Therefore, we run power flow study $C$ times by randomly sampling PV and nodal load profiles between 9 a.m. and 8 p.m.. After each run, record the highest and the lowest voltage nodes inside each IBR and VR control zone as the critical nodes for each zone.

After all $C$ power flow cases are completed, divide the critical nodes obtained into a high-occurrence group, $X$, and a low-occurrence group, $Y$, based on the occurrence threshold, $TH$, so we have

$$X = [x_1, \dots, x_u, \dots, x_U]$$

$$s.t.\ Prob(x_u) = \frac{c_{x_u}}{C} \times 100\% \geq TH \quad (2)$$

$$Y = [y_1, \dots, y_w, \dots, y_W]$$

$$s.t.\ Prob(y_w) = \frac{c_{y_w}}{C} \times 100\% < TH \quad (3)$$

where $U$ and $W$ is the total number of the high- and low-occurrence critical nodes, respectively, $c_{x_u}$ and $c_{y_w}$ is the number of times that nodes $x_u$ and $y_w$ were recorded as critical nodes, respectively.

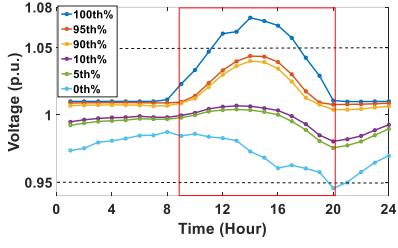

Fig. 5. Hourly nodal voltage distribution.

To determine the final set of the critical nodes for each IBR and VR control zone, the following two rules are used. First, all nodes in $X$ are considered as critical nodes. Second, a voltage difference test is conducted to decide if a node in $Y$ is a critical node. Assume there are $I_w$ times in $C$ where $y_w$ was identified as a critical node and store the index of those power flow runs in $L_w$. We can calculate the voltage difference between node $y_w$ and all the existing critical nodes in $X$ for all $I_w$ runs in $L_w$, $\Delta V_{y_w}$, as

$$\Delta V_{y_w} = \begin{bmatrix} V_{y_w}^{L_w\{1\}} - V_{x_1}^{L_w\{1\}} & \cdots & V_{y_w}^{L_w\{i_w\}} - V_{x_1}^{L_w\{i_w\}} & \cdots & V_{y_w}^{L_w\{I_w\}} - V_{x_1}^{L_w\{I_w\}} \\ \vdots & & \vdots & & \vdots \\ V_{y_w}^{L_w\{1\}} - V_{x_u}^{L_w\{1\}} & \cdots & V_{y_w}^{L_w\{i_w\}} - V_{x_u}^{L_w\{i_w\}} & \cdots & V_{y_w}^{L_w\{I_w\}} - V_{x_u}^{L_w\{I_w\}} \\ \vdots & & \vdots & & \vdots \\ V_{y_w}^{L_w\{1\}} - V_{x_U}^{L_w\{1\}} & \cdots & V_{y_w}^{L_w\{i_w\}} - V_{x_U}^{L_w\{i_w\}} & \cdots & V_{y_w}^{L_w\{I_w\}} - V_{x_U}^{L_w\{I_w\}} \end{bmatrix} \quad (4)$$

First, for the $u^{\text{th}}$ critical node in $X$, $x_u$, calculate the maximum voltage difference between $y_w$ and $x_u$ in all $I_w$ runs, $\Delta V_{y_w,X}^{max}$, as

$$\Delta V_{y_w,X}^{max}(u) = max\left(\Delta V_{y_w}(u, 1: I_w)\right) \quad (5)$$

Then, calculate the smallest $\Delta V_{y_w,X}^{max}$, $\Delta V_{y_w}^{min}$, as

$$\Delta V_{y_w}^{min} = min(\Delta V_{y_w,X}^{max}) \quad (6)$$

Let $\Delta V_{th}$ be the voltage threshold that determines whether or not a less-frequently occurred critical nodes should be included to the critical load list. If $\Delta V_{y_w}^{min} < \Delta V_{th}$, node $y_w$ will not be considered in the final critical node set. Otherwise, it will.

Note that voltage difference test is conducted to assess the voltage difference between a low-occurrence critical node and the high-occurrence critical nodes in the same IBR or VR zone. If the voltage difference is minimal, then there is no need to include $y$ as a new critical node. This will significantly reduce the number of critical nodes and improve the computational speed in practice.

As an illustration, assume $\Delta V_{th} = 0.001\ p.u.$ and there are two high-occurrence critical nodes and one low-occurrence node (with only 4 occurrences). By the first rule, the two high-occurrence critical nodes are considered as critical nodes. Assume that the voltage differences between the low-occurrence node and the two critical nodes are

$$\Delta V_1 = \begin{bmatrix} 0.0018 & 0.0018 & 0.0017 & 0.0017 \\ 4.1 \times 10^{-4} & 3.8 \times 10^{-4} & 4.5 \times 10^{-4} & 4.4 \times 10^{-4} \end{bmatrix}.$$

Then, we have

$$\Delta V_{1,X}^{max} = \begin{bmatrix} 0.0018 \\ 4.5 \times 10^{-4} \end{bmatrix}$$

$$\Delta V_1^{min} = 4.5 \times 10^{-4} < \Delta V_{th} = 0.001\ p.u.$$

Thus, this low-occurrence node will not be considered as a critical node. The flow chat for the critical node identification process is presented in Fig. 6.

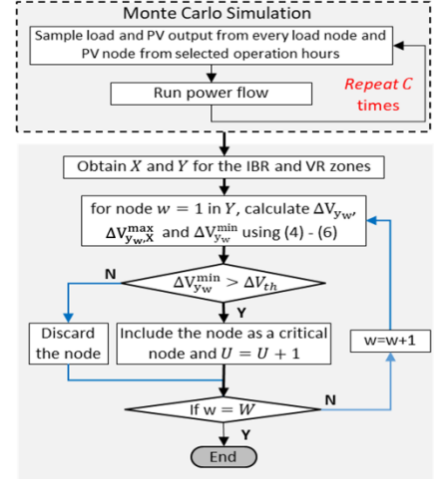

Fig. 6. Critical nodes selection process flowchart.

### C. First-stage IBR-based Zonal VVC

To implement the IBR-based zonal VVC, we need to partition the feeder into non-overlapping, weakly-coupled voltage control zones using nodal voltage correlations. Specifically, the distribution feeder is divided into $K$ PV smart inverters zones (IBR zones) per each phase based on voltage reactive power sensitivity correlations using method introduced in [21]. Then, a distributed, priority-list based zonal control scheme is developed to correct voltage violations based on voltage-load sensitivity. Note that this is a non-optimization based approach. Compared with approach introduced in [21], in this paper, only voltage measurements from critical nodes are used instead of using voltage measurements from all nodes inside the IBR zone.

*1) Clustering Method for IBR Zone Identification*

A fast incremental clustering (FIC) method is used to cluster load nodes based on voltage-load-sensitivity matrix (VLSM) with respect to reactive power changes, $VLSM_Q$. First, group loads on the same phase together so that each phase can be treated separately to account for unbalanced loads in distribution systems. Next, a power flow based method introduced in [22] is used to calculate $VLSM_Q$ as

$$VLSM_Q = \begin{bmatrix} q_{1,1} & q_{1,2} & \cdots & q_{1,N} \\ q_{2,1} & q_{2,2} & \cdots & q_{2,N} \\ \vdots & \vdots & \ddots & \vdots \\ q_{N,1} & q_{N,2} & \cdots & q_{N,N} \end{bmatrix} \quad (7)$$

$$VLSM_Q(:,j) = \begin{bmatrix} q_{1,j} \\ \vdots \\ q_{N,j} \end{bmatrix} = \frac{V_{i,j}^Q - V_i^{BASE}}{\Delta Q_j} \quad \forall\ i \in N, \forall\ j \in N \quad (8)$$

where $q_{i,j}$ is an entry in $VLSM_Q$; $N$ is the number of load nodes; $V_i^{BASE}$ is the base voltage at load node $i$ for a given scenario; $\Delta Q_j$ is the reactive power perturbation at node $j$; $V_{i,j}^Q$ is the voltage at node $i$ when perturbing node $j$ by $\Delta Q_j$.

Compute $VLSM_Q$ for 1000 scenarios and calculate the mean and the standard deviation of the $j^{\text{th}}$ column of $VLSM_Q$, as

$\mu_j$ and $\sigma_j$. Then, compute the voltage sensitivity correlation matrix, $\mathbf{C_Q}$ as

$$\mathbf{C_Q} = \begin{bmatrix} 1 & c_{1,2} & \cdots & c_{1,N} \\ c_{2,1} & 1 & \cdots & c_{2,N} \\ \vdots & \vdots & \ddots & \vdots \\ c_{N,1} & c_{N,2} & \cdots & 1 \end{bmatrix} \quad (9)$$

$$c_{m,n} = \frac{1}{N}\sum_{i=1}^{N}\left(\frac{q_{i,m}-\mu_m}{\sigma_m}\right)\left(\frac{q_{i,n}-\mu_n}{\sigma_n}\right) \quad \forall m \in N, \forall n \in N \quad (10)$$

where $c_{m,n}$ is the Pearson correlation coefficient [23] for evaluating the correlation between any two columns, $m$ and $n$, in $VLSM_Q$.

The mean voltage sensitivity correlation coefficient between a new load node $l$ and load node $v$ in an existing IBR zone $k$, $mcc_{l,k}$, is calculated as

$$mcc_{l,k} = \frac{1}{m_k}\cdot\sum_{v=1}^{m_k} c_{l,v} \quad \forall v \in m_k, \forall k \in K \quad (11)$$

where $v$ is the index of the load node inside IBR zone $k$, $m_k$ is the total number of load nodes inside the IBR zone $k$, and $K$ is the total number of IBR zones.

If $mcc_{l,k}$ calculated for all $K$ zones are below the correlation threshold, $\alpha$, a new IBR zone will be created. Otherwise, put node $l$ into an existing IBR zone with the maximum $mcc_{l,k}$. Because FIC groups load nodes by comparing $mcc_{l,k}$ to $\alpha$, increasing $\alpha$ will increase the number of IBR zones. The FIC algorithm is summarized in Algorithm 1.

| Algorithm 1: Fast Incremental Clustering | |
|---|---|
| 1: | Set the number of IBR zones to be $K = 1$ and select a correlation threshold, $\alpha$. |
| 2: | Place node 1 into IBR zone 1. |
| 3: | Compute the correlation matrix $C_Q$ using (9). |
| 4: | **for** node $l=2...N$ |
| 5: | Use (11) to calculate $mcc_{l,k}$ for all $K$ IBR zones. |
| 6: | **if** $\max_{k \in K}(mcc_{l,k}) \geq \alpha$ |
| 7: | Place $l$ in the zone having maximum $mcc_{l,k}$. |
| 8: | **else** |
| 9: | $K = K + 1$ |
| 10: | Place $l$ in the new IBR zone $K$. |
| 11: | **end** |
| 12: | $l = l + 1$ |
| 13: | **end** |

*2) IBR Zonal Control Mechanism*

In real-time operation, at time $t$, voltage measurements $\widehat{V}_{k,u}(t)$ ($\forall u \in U$) at critical nodes in each IBR zone are sent to the IBR controller. If a voltage violation is detected in the $k^{\text{th}}$ IBR zone (See point 1 in Fig. 7), the $k^{\text{th}}$ IBR zone controller will identify the *extreme critical* node, $r$, which is the node with the highest over-voltage, or the node with the lowest under-voltage. The *extreme voltage violation*, $\Delta V_{k,r}(t)$ is calculated as

$$\Delta V_{k,r}(t) = \begin{cases} \max_{u \in U}\widehat{V}_{k,u} - V_{1\_max} & if \max_{u \in U}\widehat{V}_{k,u} \geq V_{1\_max} \\ V_{1\_min} - \min_{u \in U}\widehat{V}_{k,u} & if \min_{u \in U}\widehat{V}_{k,u} \leq V_{1\_min} \end{cases} \quad (12)$$

where $V_{1\_max}$ and $V_{1\_min}$ is the predetermined first-stage nodal voltage upper and lower limit, respectively.

The reactive power, $\Delta Q_{k,Req}(t)$, for correcting $\Delta V_{k,r}(t)$ is calculated as

$$\Delta Q_{k,Req}(t) = \frac{\Delta V_{k,r}(t)}{\alpha \cdot q_{r,r}} \quad (13)$$

where $q_{r,r}$ is the voltage-reactive power sensitivity at critical node $r$ from the $VLSM_Q$.

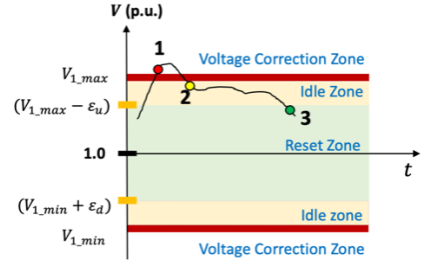

Fig.7. IBR-based voltage correction process.

To regulate the voltage at critical node $r$ in the $k^{\text{th}}$ IBR zone, a distributed, priority-list based zonal IBR control scheme is developed. Let $PL_k$ represent the priority list in the in $k^{\text{th}}$ IBR zone. All node $v$ in IBR zone $k$ will be assigned a priority, $PL_{k,v}$, which is calculated from $c_{v,r}$ using (10) and ranked in the descending order. The controller dispatch IBRs starting from the top of the priority list in a sequential manner until $\Delta Q_{k,Req}(t)$ is met or all IBR resources are depleted.

To avoid voltage oscillations, a voltage return band, $[(V_{min,1} + \varepsilon_d),(V_{max,1} - \varepsilon_u)]$, is defined, where $\varepsilon_u$ and $\varepsilon_d$ are the two voltage return margins. If voltage at the extreme critical node returns to the idle zone (see point 2 in Fig. 7), the IBR controller will do nothing; if the voltage returns to the reset zone (see point 3 in Fig. 7), the IBR controller will return IBRs to the unity power factor mode based on the reversed priority list where (12) and (13) are used to calculate $\Delta Q_{k,Req}(t)$. This will let the IBRs return to the unity power factor operation model based on the reversed priority list.

Note that in this method, only voltages measurements at the critical nodes in an IBR zone are used as the inputs to the IBR controller, while for the method introduced in [21], voltage measurements from all nodes in the IBR zone are required.

*D. Second-Stage VR-based Zonal VVR*

In the second-stage operation, at time $T$, each VR controller $g$ ($\forall g \in G$) monitors its secondary voltage, $V_{g,2}(T)$, and voltage at the critical nodes, $\widehat{V}_{g,u}(T)$ ($\forall u \in U$) in its zone. Let $V_{2\_max}$ and $V_{2\_min}$ be the predetermined second-stage nodal voltage upper and lower limits, respectively. The *extreme voltage change* at time $T$, $\Delta V_{g,u}(T)$ be calculated as

$$\Delta V_{g,u}(T) = \begin{cases} \max_{u \in U}\left(\widehat{V}_{g,u}(T)\right) - V_{g,2}(T) & if \max_{u \in U}\left(\widehat{V}_{g,u}(T)\right) \geq V_{2\_max} \\ V_{g,2}(T) - \min_{u \in U}\left(\widehat{V}_{g,u}(T)\right) & if \min_{u \in U}\left(\widehat{V}_{g,u}(T)\right) \leq V_{2\_min} \end{cases} \quad (14)$$

Then, the VR setpoint at time $T$, $V_{set}^g(T)$ is calculated as

$$V_{set}^g = \begin{cases} \operatorname{mean}_{u \in U}\widehat{V}_{g,u}(T), & if \max_{u \in U}\left(\widehat{V}_{g,u}(T)\right) < V_{2\_max} \\ & and \min_{u \in U}(\widehat{V}_{g,u}(T)) > V_{2\_min} \\ V_{VR,2}^g - \max\left(\Delta V_{g,u}(T),\frac{DB_g}{2}\right) & if \max_{u \in U}\left(\widehat{V}_{g,u}(T)\right) \geq V_{2\_max} \\ V_{VR,2}^g + \max\left(\Delta V_{g,u}(T),\frac{DB_g}{2}\right) & if \min_{u \in U}\left(\widehat{V}_{g,u}(T)\right) \leq V_{2\_min} \end{cases} \quad (15)$$

In (15), $\Delta V_{g,u}(T)$ is compared with $\frac{DB_g}{2}$ to ensure the voltage set point is adjusted to compensate for the extreme voltage change. As illustrated in Fig. 8, at time $T_1$, the VR controller observes a voltage rise. Because there is no voltage violations in the VR zone, the VR controller will assign a new setpoint, $V_{set}^g(T_1)$, to avoid an unnecessary tap change. At time $T_2$, the maximum voltage at critical nodes exceeds $V_{2\_max}$, the controller will calculate $V_{set}^g(T_2)$ using (15), consequently lead to a tap change in order to bring $V_{g,2}(T_2)$ back within $V_{set}^g \pm \frac{DB^g}{2}$. Note that a time delay, $t_d^g$, is used when executing the VR tap change.

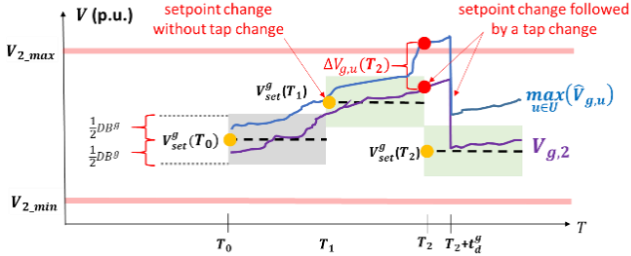

Fig. 8. VR-based zonal VVR strategy.

To coordinate the voltage regulation between VR and IBR controllers, we set

$$V_{max,1} = V_{max,2} - \Delta V \quad (17)$$

where $\Delta V$ is a voltage margin. Also, let $t < T$ so that the IBR control will be implemented in a faster pace. Those settings allows the IBR controllers to take the first-stage actions for correcting the local voltage excursions while the VR controllers will correct large, system-wise voltage deviations.

## IV. SIMULATION RESULTS

A 12.47 kV three-phase unbalanced actual distribution feeder model representing a real circuit located in North Carolina is selected to verify the effectiveness of the proposed methods. The topology of the feeder is shown in Fig. 3. The feeder includes 1283 nodes, 370 of which are load nodes. The feeder load disaggregation algorithm presented in [20] has been used to allocate 1-min resolution residential load profiles from Pecan Street [24] to every load node on a test feeder. This step is important to give each load node a realistic individual load profile and will prepare the nodal load profiles for running quasi-static power flow simulations. The disaggregation algorithm was applied to the distribution feeder and it has allocated 427 households aggregated to a substation load peak at 1.82 MW. To model high-PV penetration scenarios, each household has a 5 kW PV system connected to it summing to a total 2.135 MW of PV installation. It is assumed that each inverter is sized such that its total apparent power can reach up to 110% of its rated active power capacity. Also, we assume that each PV system is equipped with a smart inverter that can inject and absorb reactive power.

### A. IBR Zone Identification

We performed FIC at multiple randomly sampled PV and nodal load profiles between 9:00 a.m. and 8:00 p.m. to observe the voltage sensitivity correlations effect on the number of generated clusters during different operation conditions. As shown in Fig. 9, load nodes on different phases (i.e., phase $a$, $b$, and $c$) are clustered using multiple correlation thresholds $\alpha$ under 1000 different PV/load scenario.

There are different numbers of IBR zones on different phases, so each phase needs to be treated separately. FIC generated a stable number of clusters up to $\alpha \leq 0.96$. When using $\alpha \geq 0.97$ to cluster load nodes, the number of nodes in a cluster may vary for different operation conditions.

This shows that the correlation between voltage and reactive power is consistent for nodes on the same phase under a wide range of operation conditions as long as we don't use an extremely high $\alpha$. In this paper, we choose $\alpha = 0.96$, which results in one IBR zone in phase $b$ and two IBR zones in phase $a$ and $c$, respectively. The IBR zones are shown in Fig. 3(a).

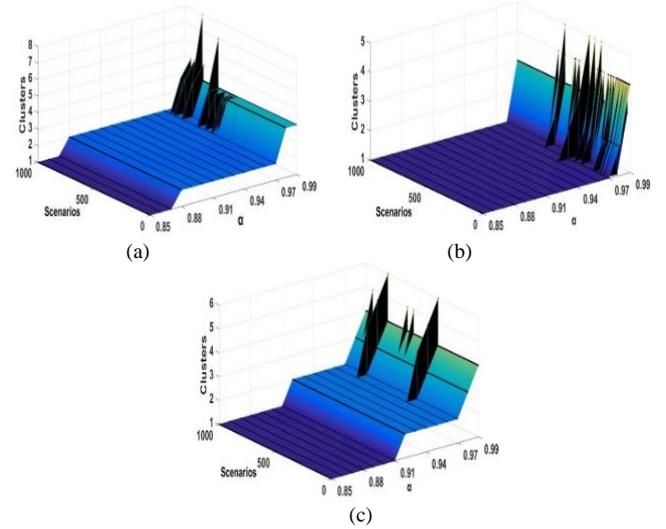

Fig. 9. Impact of the correlation threshold, $\alpha$, on the number of clusters for (a) Phase $a$, (b) Phase $b$, and (c) Phase $c$.

### B. Critical Nodes Selection

Five IBR zones (see Fig. 3a) were identified when setting $\alpha = 0.96$. The utility-owned assets include an OLTC at the feeder head, one three-phase SVR and two single-phase SVRs, resulting in four VR zones (see Fig. 3b). The OLTC/SVR are all single-phase auto-transformers.

Run Monte Carlo simulation for 10,000 between 9:00 a.m. and 8:00 p.m. (see Fig. 5) and set $TH = 5\%$ and $V_{th} = 0.001$ p.u.. The numbers of critical nodes identified for the IBR and VR zones are presented in Table I and Table II, respectively.

A sensitivity study is performed to quantify the impact of the number of Monte Carlo run on the number of identified critical nodes. As shown in Fig. 10, the total number of identified critical nodes for the IBR and VR zones stabilized after 30,000 run. Because the VR control zones overlap with the IBR zones (see Fig. 3), the critical nodes in the IBR zones are a subset of the critical nodes in the VR zones. If we count non-overlapping critical nodes in both the IBR and VR zones together, there are 51 critical nodes identified, representing 3.9% of the nodes on this feeder. In many cases, infrequently-occurred critical nodes can be replaced by frequently-occurred critical nodes. Therefore, performing a voltage difference check can significantly reduce the number of critical nodes.

TABLE I
IBR Zones Critical Nodes

| IBR Zone | X | Y | Final Set of Critical Nodes |
|---|---|---|---|
| 1 | 4 | 16 | 4 |
| 2 | 3 | 15 | 3 |
| 3 | 4 | 32 | 8 |
| 4 | 4 | 7 | 6 |
| 5 | 2 | 32 | 11 |
| Total nodes | 17 | 102 | 32 |

TABLE II
VR Zones Critical Nodes

| VR Zones | Regulators | X | Y | Final Set |
|---|---|---|---|---|
| 1 | 1.a | 3 | 7 | 3 |
|   | 1.b | 5 | 8 | 5 |
|   | 1.c | 4 | 7 | 6 |
| 2 | 2.a | 4 | 6 | 4 |
| 3 | 3.a | 3 | 15 | 3 |
|   | 3.b | 4 | 30 | 11 |
|   | 3.c | 2 | 32 | 11 |
| 4 | 4.a | 3 | 7 | 3 |
| Total nodes |   | 28 | 112 | 46 |

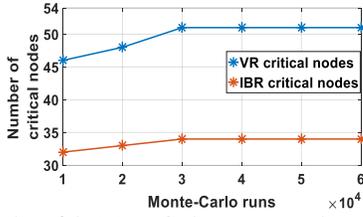

Fig. 10. Number of the Monte Carlo runs versus the number of critical nodes identified

### C. Performance of the Coordinated Voltage Control

To test the performance of the proposed 2-stage, coordinative VVC introduced in Section III, the following metrics are used: the number of voltage violations, $N_V^{TOT}$, the total number of tap changes of all VR devices, $N_{VR}^{TOT}$, and the total reactive power output of all IBR inverters, $Q_{IBR}^{TOT}$.

In the base case, only the utility-owned VR devices are used and all OLTC/SVR are operating under local control with $V_{set}=123$ V, $DB = 4$ V, and time delays between 30 s to 60 s.

In the coordinative voltage control cases, set $V_{2\_max} = 1.049\ p.u.$ and $V_{2\_min} = 0.96\ p.u.$ so the zonal voltage will be controlled to be within the ANSI limits. Let the IBRs adjust reactive power output every one minute (i.e., $t = 1$) with $\varepsilon_u = 0.001\ p.u.$ [21] and let VR devices adjust their voltage setpoints every two minutes (i.e., $T = 2$) with $DB = 4$V. Set the time delays of VR devices as: $TD_{OLTC} = 75$ s, $TD_{SVR2} = TD_{SVR3} = 90$ s, $TD_{SVR4} = 105$ s. To evaluate the impact of $\Delta V$ (voltage margin introduced in (17)) on voltage coordination between the VR device and IBRs, we tested four $\Delta V$ settings: 0, 0.002, 0.004, and 0.006 V.

The results of a winter month, where voltage regulations have been frequently observed, are presented to demonstrate the performance of the proposed algorithm. The simulation time step is 1 minute. As shown in Table III, if the IBRs do not provide voltage regulation, although the voltage regulators act frequently, voltage violations still persist. This can also be seen in the voltage profile shown in Fig. 11 (a).

The results in Table III also show that a larger $\Delta V$ can significantly reduce the tap changes but increase the reactive power outputs of the IBRs. This is because by setting up a larger $\Delta V$, we shift more voltage regulation tasks from the VR devices to the IBRs so that voltage fluctuations are mainly taken care of by the local IBR devices.

Figure 11(b) shows that voltages at all nodes on the feeder are maintained within the ANSI limits. Note that this is achieved by monitoring only the voltages at the critical nodes.

Also, the proposed first-stage control (i.e. reactive power compensation from IBRs) is capable of estimating the required reactive power support from PV smart inverters when voltage violations exist and reset the smart inverters back to unity power factor when the voltages return within proper limits, as shown in Fig. 12. The total reactive power contribution by each IBR zone inverters for a winter month is shown in Fig. 13(a).

Note that in (13), the voltage-reactive power sensitivity, $q_{r,r}$, at critical node $r$ is used to estimate the reactive power required from each IBR. In general, the voltage sensitivities vary when the feeder is reconfigured or the nodal load pattern changes. For the feeder used in this simulation, the voltage sensitivity variations between a winter day (low load and high PV) and a summer day (high load and high PV) are typically within 5%. In practice, we recommend that the voltage sensitivities be calculated offline to account for seasonality load changes and feeder reconfiguration events.

TABLE III
Simulation Results Summary for a Winter Month

| Cases | $N_V^{TOT}$ | $N_{VR}^{TOT}$ | $Q_{IBR}^{TOT}$ (kvarh) |
|---|---|---|---|
| Base case | 769,182 | 1157 | N.A. |
| Proposed $\Delta V = 0$ | 708 | 388 | -4,685 |
| Proposed $\Delta V = 0.002$ | 560 | 291 | -7,423 |
| Proposed $\Delta V = 0.004$ | 450 | 271 | -9,583 |
| Proposed $\Delta V = 0.006$ | 149 | 204 | -17,180 |

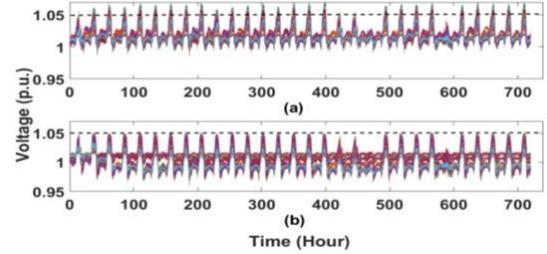

Fig. 11. Comparison of the nodal voltage profiles. (a) The base case, and (b) the 2-stage VVC case with $\Delta V = 0.006\ p.u.$

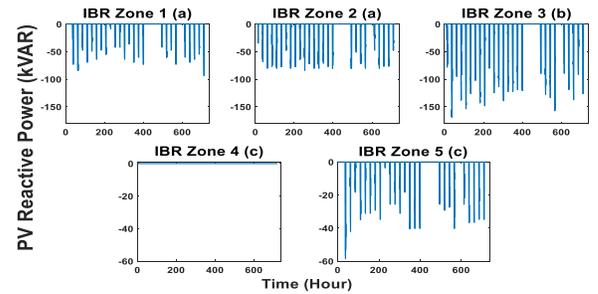

Fig.12. Reactive power support from IBR zones

As shown in Figs. 14 and 13(b), the number of tap changes is reduced significantly. This demonstrates the effectiveness of the IBR and VR coordination scheme as well as the online voltage setpoint tuning method.

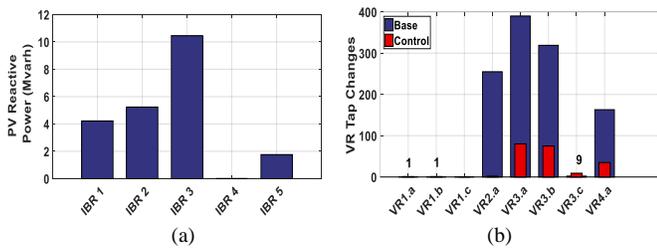

Fig.13. (a) Reactive power generation in each IBR zone and (b) VRs tap changes in each VR zone (base cases versus controlled cases)

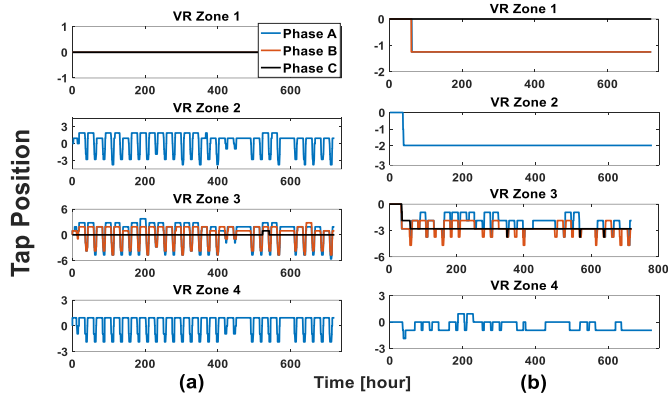

Fig.14. Tap change operations for (a) the base case, (b) the 2-stage VVC case with $\Delta V = 0.006\ p.u.$

## V. CONCLUSION

In conclusion, the proposed two-stage, zonal Volt/VAR control scheme for coordinating IBRs with utility-owned VR devices can effectively reduce the number of tap changes of the VR devices while maintaining nodal voltages within the operation limit. Because the IBR and VR control zones are partitioned based on the nodal voltage-reactive power sensitivities per each phase, the unbalanced loads can be accounted for when regulating voltages. Coordination between IBRs and VR devices can be successfully achieved by selecting different voltage control targets and letting the IBR control execute at a smaller time interval than the VR control. Simulation results demonstrated that this enables IBRs to react to very fast, small voltage fluctuations while letting the VRs correct persistent, larger voltage deviations. Finally, the proposed online voltage setpoint tuning mechanism significantly reduced the number of tap changes. The algorithm uses real-time voltage measurements from only the critical nodes (less than 3.9% of total nodes) and requires neither optimization nor the feeder model, making the computing time negligible with minimal sensing and communication needs. Thus, the algorithm can be deployed on grid edge computation platforms without retrofitting the existing VR devices.